# gSketch: On Query Estimation in Graph Streams

Peixiang Zhao
University of Illinois
Urbana, IL 61801, USA
pzhao4@illinois.edu

Charu C. Aggarwal
IBM T. J. Watson Res. Ctr.
Hawthorne, NY 10532, USA
charu@us.ibm.com

Min Wang
HP Labs, China
Beijing 100084, P. R. China
min.wang6@hp.com

## ABSTRACT

Many dynamic applications are built upon large network infrastructures, such as social networks, communication networks, biological networks and the Web. Such applications create data that can be naturally modeled as *graph streams*, in which edges of the underlying graph are received and updated sequentially in a form of a stream. It is often necessary and important to summarize the behavior of graph streams in order to enable effective query processing. However, the sheer size and dynamic nature of graph streams present an enormous challenge to existing graph management techniques. In this paper, we propose a new graph sketch method, gSketch, which combines well studied synopses for traditional data streams with a sketch partitioning technique, to estimate and optimize the responses to basic queries on graph streams. We consider two different scenarios for query estimation: (1) A graph stream sample is available; (2) Both a graph stream sample and a query workload sample are available. Algorithms for different scenarios are designed respectively by partitioning a global sketch to a group of localized sketches in order to optimize the query estimation accuracy. We perform extensive experimental studies on both real and synthetic data sets and demonstrate the power and robustness of gSketch in comparison with the state-of-the-art global sketch method.

## 1. INTRODUCTION

Recent scientific and technological advances have resulted in a proliferation of graph-structured data, such as E-R schemas, chemical compounds, biological or social networks, work flows and the Web. Accordingly, many data management and mining applications have been extensively studied in the graph domain [2, 4, 12]. However, much of the focus of past research has been on a (or a set of) static graph(s) of relatively modest size. In recent years, numerous network applications have witnessed streams being defined over the massive graph infrastructure [15, 19, 20, 16, 3], in which the entire graph of interest is no longer available all the time, but individual edges of the graph are received and updated rapidly over time in a form of a stream. These newly emerging graphs are referred to as *graph streams*. Some noteworthy examples of graph streams correspond to the activities overlaid on the Web graphs [28], social networks and communication networks. In these cases, the vertices of the graph correspond to different web pages, actors, or IP addresses and the edges represent the links or communication activities among them. Such graph streams may change rapidly in the context of a massive domain of potential edges.

A key property of a graph stream is that it is dynamically updated and the speed of incoming edges can be very high. Therefore, the standard stream constraint of being able to process every edge only once applies naturally in this scenario. Besides its dynamic nature, an additional challenge arises due to the massiveness of a graph stream. In theory, the number of distinct edges of a graph stream is quadratically related to the number of vertices, and thus it becomes prohibitive to manage such a huge number of edges explicitly. For example, in a social network containing $10^7$ users, the number of distinct edges is of the order of $10^{14}$. The number of interactions, such as the instant messages being sent among individuals at the moment, is even prohibitively larger. Without an efficient storage of the underlying graph, it becomes almost impossible to enable effective query processing on graph streams.

In this paper, we consider some typical queries relevant to graph streams:

- **Edge Query:** We estimate the frequency of particular edges in a graph stream;

- **Aggregate Subgraph Query:** We determine the aggregate frequency behavior of constituent edges of a subgraph.

Despite being primitive, both edge query and aggregate subgraph query are nontrivial in graph streams, and they serve as the building bricks of many advanced querying and mining operations of graph streams [6, 20, 16]. Some applications of such queries are as follows:

1. In social networking applications, vertices represent the participants of a social network, and edges correspond to interactions among the participants. For very large and frequently updated social networks, the underlying graph streams have a rapid rate of incoming edges. An edge query is to estimate the communication frequency between two specific friends, while an aggregate subgraph query is to estimate the overall communication frequencies within a community;

2. In network intrusion applications, vertices of the graph stream represent distinct IP addresses, and edges correspond to network attacks between IP pairs. IP attack packages involving different IP pairs are received dynamically in a stream fashion. An edge query is to estimate the attack frequency between a pair of IP addresses, and an aggregate subgraph query is to estimate the overall attack frequencies of a subset of interconnected IP addresses.

It is possible to design a straightforward solution by building a global synopsis structure corresponding to the entire graph stream for query estimation. Note here any well studied sketch method [5, 23, 14, 11] can be leveraged, which, however, is blind to the underlying structural behavior of graph streams. In real applications, the frequencies of edges are often extremely skewed over different regions of the underlying graph. It is therefore inefficient to use a single global sketch for the whole graph stream, which does not consider such structure-related skewness to its advantage.

In this paper, we propose a new graph sketch method, gSketch, for querying large graph streams. An important contribution of gSketch is that it resolves the challenges of query estimation *by making use of typical local and global structural behavior of graph streams*. We consider well known properties of graph streams in real applications, such as the global heterogeneity and local similarity, in conjunction with coarse and easy to compute *vertex-specific* statistics, to create an intelligent partitioning of the virtual global sketch toward optimizing the overall query estimation accuracy. In this way, incoming queries can be answered by the corresponding partitioned local sketches, upon which the query estimation accuracy can be improved. The motivation to use the vertex-specific statistics of the graph stream during sketch partitioning is twofold. First, the locality similarity within the vicinity of vertices can be effectively encoded and leveraged during sketch partitioning. Second, although the number of potential edges may be too large to be characterized, the number of vertices of a graph stream is often much more modest [19], and the vertex-based statistics can be easily quantified during query estimation. We consider two practical scenarios for sketch partitioning: (1) a graph stream sample is available, and (2) both a stream sample and a query workload sample are available. Efficient sketch partitioning algorithms under different scenarios are designed respectively and our experimental results on both real and synthetic graph streams have demonstrated the effectiveness and efficiency of gSketch. In both scenarios, gSketch achieves up to an order of magnitude improvement in terms of the query estimation accuracy, compared with the state-of-the-art global sketch method.

The remainder of this paper is organized as follows. We discuss related work in Section 2. In Section 3, we introduce a general framework for querying graph streams. We will have a broad discussion on how sketches can be used for query estimation in graph streams, and the potential problems of a direct application of a global sketch structure. In Section 4, we are focused on the issues and algorithms of sketch partitioning under different scenarios, which are determined by the availability of different sample data. Query processing in the presence of a group of partitioned sketches is detailed in Section 5. Section 6 contains our experimental studies, and Section 7 concludes the paper.

## 2. RELATED WORK

The problem of querying and mining data streams has been studied extensively [25, 1] in recent years. The earliest work in the graph domain, however, was proposed in [21]. The subsequent work explored methods for counting the number of triangles [6, 8], determining shortest paths [19], estimating PageRank scores [16], mining dense structural patterns [3], and characterizing the distinct degree counts of the nodes in the multi-graph scenario [15]. An excellent survey on mining and querying graph streams can be found as well [24]. Surprisingly, none of the previous work has focused on the query estimation issue on large graph streams.

On the other hand, sketch synopses, including but not limited to AMS [5], Lossy Counting [23], CountMin [14] and Bottom-$k$ [11], have proven to be effective data structures in the general stream scenario. Such sketches however may not work well for the case of graph data. For example, they do not consider the underlying correlations of the edge frequencies in the graph stream. That is, only partial information available in graph streams is leveraged in these sketch-based structures. As a result, the sketching methods have to be re-examined and designed specifically to accommodate the new challenges and characteristics inherent in graph streams.

Although the sketch partitioning technique has been proposed in the general data stream domain for join size estimation [17], it is in the context of non-structural data. From an algorithmic point of view, this method works with the AMS method as a base, and attempts to minimize the variance of attribute values within each partition. This is based on how the error of join-size estimation is computed, in which the weighted sum of the variances of join attributes needs to be minimized. This approach is quite different from our sketch partitioning method, gSketch, on graph streams. In gSketch, we make use of the structural frequency behavior of vertices in relation to the edges for sketch partitioning. In other words, the structural nature of a graph stream makes it quite different from the former sketch-partitioning problem, which has applied to the multi-dimensional data.

## 3. THE ALGORITHMIC FRAMEWORK

In this section, we will discuss an algorithmic framework for query estimation in graphs streams. We first formulate the problem of query estimation for the case of graph streams. A straightforward solution is then proposed to construct a global sketch for the entire graph stream. The main limitation of this global sketch method is that the structural properties of graph streams are totally ignored during query estimation. Such limitation also motivates us to consider leveraging the underlying structural properties of graph streams, thus resulting in our sketch-partitioning based solution, gSketch.

### 3.1 Problem Definition

Given a graph stream, we assume its underlying graph $\mathcal{G}$ can be defined as $\mathcal{G} = (\mathcal{V}, \mathcal{E})$, where $\mathcal{V}$ is a vertex set of $\mathcal{G}$. For each vertex $x \in \mathcal{V}$, there is a string $l(x)$ attached to $x$ as the label of $x$. $\mathcal{E}$ is the edge set of $\mathcal{G}$ and every edge $(x, y) \in \mathcal{E}$ is a directed edge[1]. The incoming graph stream contains elements $(x_1, y_1; t_1), (x_2, y_2; t_2), \ldots (x_i, y_i; t_i) \ldots \ldots$, where the edge $(x_i, y_i)$ is encountered at the time-stamp $t_i$. In some applications, a frequency $f(x_i, y_i, t_i)$ is associated with the edge $(x_i, y_i)$ at $t_i$. For example, in a telecommunication application, the frequency $f(x_i, y_i, t_i)$ may denote the number of seconds in a phone conversation from a person $x_i$ to another person $y_i$ starting at the time-stamp $t_i$. If not specified explicitly, we assume $f(x_i, y_i, t_i) = 1$ by default.

---
[1] In the event of an undirected graph, lexicographic ordering on vertex labels can be used in order to determine the direction of the edge. Ties are broken arbitrarily.

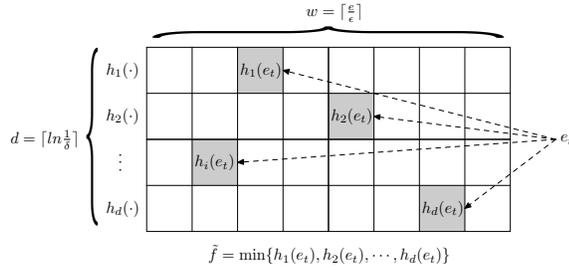

**Figure 1: A CountMin Sketch**

Hitherto, the graph stream as defined above has been frequently encountered in a number of application domains such as network intrusion detection, telecommunication, and social networks. Some representative queries in such graph streams can be formulated as follows:

- **Edge Query:** To estimate the overall frequency of the edge $(x, y)$: $\tilde{f}(x, y) = \sum_{t_i \in \mathcal{T}} f(x, y, t_i)$, where $\mathcal{T}$ can be the lifetime of the graph stream or a specific time window of interest.

- **Aggregate Subgraph Query:** To estimate the aggregate frequency of the constituent edges of a subgraph $G = \{(x_1, y_1), \ldots, (x_k, y_k)\}$: $\tilde{f}(G) = \Gamma(\tilde{f}(x_1, y_1), \ldots, \tilde{f}(x_k, y_k))$, where $\Gamma(\cdot)$ is an aggregate function of interest, such as MIN$(\cdot)$ or AVERAGE$(\cdot)$. For example, when $\Gamma(\cdot) = $ SUM$(\cdot)$, it can summarize the total frequency of all the edges of the subgraph $G$, i.e., $\tilde{f}(G) = \sum_{i=1}^{k} \tilde{f}(x_i, y_i)$.

Aggregate subgraph query is essentially a derivative of edge query in the sense that it is performed on a bag of edges belonging to a subgraph, so it can be naturally resolved by the use of the function $\Gamma(\cdot)$ on all query results of the constituent edge queries. Therefore, we will mostly be focused on edge query estimation. The generalization from edge query towards aggregate subgraph query is straightforward and will be discussed further in Section 6.

### 3.2 A Global Sketch Solution

In this section, we will discuss a straightforward solution, denoted as Global Sketch, for query estimation in graph streams. Global Sketch is an intuitive application of any sketch method [5, 23, 14, 11, 13] for summarizing the entire graph stream, provided that the edges of the graph stream can be represented and accommodated appropriately in the traditional data stream scenario. Our discussion will focus on one specific sketch method, CountMin [14], while our analysis below can be easily generalized toward other sketch methods in an analogous way.

A CountMin sketch consists of a 2-dimensional array with a width of $w = \lceil e/\epsilon \rceil$ and a depth of $d = \lceil \ln(1/\delta) \rceil$, and thus there exist $w \cdot d$ cells in total in the sketch. Here $e$ is the base of the natural logarithm. $\epsilon$ and $\delta$ are user-specified parameters, which imply that the error of answering a query is within a factor of $1 + \epsilon$ of the true value with probability at least $1 - \delta$. In the CountMin sketch, we select $d$ pairwise independent hash functions, $h_1, \ldots, h_d$. Each $h_i$ ($1 \leq i \leq d$) uniformly maps onto random integers in the range $[0, w-1]$ and corresponds to one of $d$ 1-dimensional arrays with $w$ cells each. These $d$ hash functions are used to update the frequency of an element from a data stream on different mapping cells in the 2-dimensional data structure. For example, consider a data stream with elements drawn from a universe of domain values. When a new element $e_t$ is received at the time-stamp $t$, we apply each of the $d$ hash functions upon $e_t$ to map onto a number in $[0 \ldots w-1]$. The value of each of $d$ cells, $h_i(e_t)$, is incremented accordingly by 1. In order to estimate the frequency of the element during query processing, we choose the set of $d$ cells onto which the $d$ hash-functions map, and determine the minimum value among all these $d$ cells as the query estimation result. An example of a CountMin sketch is illustrated in Figure 1.

Theoretically, if $f$ is the true frequency value of the element $e_t$, the estimated frequency, $\tilde{f}$, can be lower bounded by $f$, because we are dealing only with non-negative frequencies, and value collisions during hashing can only cause overestimation. A probabilistic upper-bound of $\tilde{f}$ can be determined as well [14]. Given a data stream with $N$ arrivals till the time-stamp $t$, the estimate $\tilde{f}$ is at most $f + e \cdot N/w$ with probability at least $1 - e^{-d}$, i.e., w.h.p.,

$$f \leq \tilde{f} \leq f + e \cdot N/w \quad (1)$$

Note that the probability of the error-bound being violated reduces exponentially with $d$. The CountMin sketch has proven to be accurate for many practical scenarios in traditional data streams [14].

Analogously, the CountMin sketch can be directly applied on graph streams for query estimation by treating each edge as an element with a unique identifier. We note that the edge $(x_i, y_i)$ can be represented as a string $l(x_i) \oplus l(y_i)$ where $\oplus$ is the concatenation operator on the vertex labels of $x_i$ and $y_i$. This string can be hashed as the key of the edge $(x_i, y_i)$ onto the CountMin sketch to maintain the frequency of $(x_i, y_i)$.

However, such an approach, Global Sketch, has proven to be ineffective in the graph stream domain. This is because when $N$ edges have been received in a CountMin sketch with hash range $w$, the (absolute) query estimation error $|f - \tilde{f}|$ is proportional to $N/w$, as shown in Equation (1). Therefore, the *relative estimation error* of an edge query with frequency $f$ is proportional to $N/(w \cdot f)$, which can be extremely large for small values of $f$ and large values of $N$. In practice, the edge frequency distribution of a graph stream can be quite uneven and those low frequency portions of the underlying graph can be very relevant for querying, and may show up repeatedly in the workload. Furthermore, the number of edges $N$ can be extremely large for graph streams. Therefore Global Sketch may generate inaccurate estimation results. Due to the natural vulnerability and limitations of Global Sketch, we need to design a more effective approach specifically for massive graph streams.

### 3.3 Broad Intuition for a Better Solution

The graph streams such as those found on the Web and various applications are typically not random. They often exhibit both local and global structural properties which are potentially useful in sketch construction and query estimation. Some common properties are characterized as follows:

- **Global Heterogeneity and Skewness:** The relative frequency distribution of different edges in a massive graph stream is nonuniform and often observed extremely uneven [18, 27];

- **Local Similarity:** Within structurally localized regions of the graph, relative frequencies of the edges are often correlated with one another [18, 7, 10]. Although this does not mean that the frequency behavior is identical within a locality, the correlations of the edge frequencies in a local region are considerable.

These empirical observations provide us with useful hints for the design of a more effective graph sketch method, denoted as gSketch, for query estimation. The key idea of gSketch is to partition a virtual global sketch corresponding to the entire graph stream to a set of smaller localized sketches, such that edges of different structural localities in the underlying graph can be mapped onto different partitioned local sketches. Therefore edges with similar frequency correlation can be updated and queried by the same local sketch during stream maintenance and query estimation. In this way, the estimation error of each local sketch is much lower compared with the case when a single global sketch is used. This ultimately helps improve the overall query estimation accuracy of gSketch over that of Global Sketch. It is worth noting that most often the data samples of a graph stream are readily available. In some other cases, the query workload samples may also be available. Therefore, it becomes possible for us to fully leverage such sample information with encoded structural properties of graph streams for effective sketch partitioning.

## 4. SKETCH PARTITIONING

In this section, we will introduce our sketch partitioning algorithms, which are performed as a pre-processing step on the sample data before the actual sketch structures are populated with the graph stream. Our goal of sketch partitioning is to maintain the graph streams with sufficient frequency uniformity within each partitioned sketch, such that the query estimation can be optimized over the entire graph stream. Each localized sketch in the partitioning is designed for summarizing the edge frequencies associated with particular source vertices. Therefore, it becomes much easier to maintain the sketch partitioning information in main memory, as the number of vertices is significantly smaller than the number of edges in the underlying graph.

As discussed earlier, we consider two sampling scenarios:

1. In the first scenario, a sample of the original graph stream is available. However, specific information about query workloads is unavailable.

2. In the second scenario, a sample of the original graph stream as well as a sample of the query workload is available. In this case, the sketch partitioning can be further optimized with the additional information of query workloads.

Before discussing the specific algorithms, we will introduce some notations and definitions. We denote the frequency of edge $(i, j)$ by $f(i, j)$, where $i, j \in \mathcal{V}$. This value is the one to be estimated during query processing, and is therefore not explicitly available. In fact, the edge frequency cannot even be explicitly stored in the case of graph streams because the number of potential edges can be exponentially large. The *relative frequency of a vertex* $i$, denoted as $f_v(i)$, is defined as the sum of the frequencies of the edges emanating from $i$, i.e.,

$$f_v(i) = \sum_j f(i,j) \quad (i,j) \in \mathcal{E} \qquad (2)$$

The *out degree* of a vertex $i$, denoted as $d(i)$, is defined as follows:

$$d(i) = \sum_j \theta(i,j) \quad (i,j) \in \mathcal{E} \qquad (3)$$

where

$$\theta(i,j) = \begin{cases} 0 & (i,j) \text{ is not in the graph stream} \\ 1 & \text{otherwise} \end{cases}$$

### 4.1 Sketch Partitioning with Data Sample

In this section, we will discuss the process of sketch partitioning in the presence of a data sample only. In order to construct the sketch partitions, we would like to group together structural regions of the graph stream with similar frequency behavior, which ultimately helps optimize the query estimation accuracy. However, since we are trying to estimate the edge frequencies to begin with, this frequency information for associating the edges with the corresponding sketch partitions is unfortunately not available directly. Therefore, it would seem that there is no practical way to ensure the regions with similar frequencies are assigned to the same partition. However, as discussed in Section 3.3, it is possible to exploit the structural properties of graph streams to efficiently approximate the frequency behavior of the edges in different structural localities and create the sketch partitioning accordingly. In order to make the analysis clearer, we first make an oracle assumption that the frequency $f(i, j)$ of the edge $(i, j)$ over the entire graph stream is known in advance. Later, we will relax this assumption by leveraging the structural characteristics of graph streams for frequency estimation.

Let us assume that there is a total of $r$ $(r \geq 1)$ partitions of the global CountMin sketch and $S_i$ is the localized sketch corresponding to the $i$th partition $(1 \leq i \leq r)$. The total space, which is essentially the available main memory, is allocated equally to each partition by evenly dividing the width of the global CountMin sketch while keeping the depth of each partitioned sketch to be the same as that of the global CountMin sketch. In other words, the width of $S_i$ is $w_i = w/r$ and the depth of $S_i$ is $d_i = d$, where $w$ and $d$ are the width and depth of the global sketch, respectively. In this way, we can ensure the same probabilistic guarantee of frequency estimation, $1 - e^d$, across all partitioned sketches, as indicated in Equation (1). Let $F(S_i)$ be the sum of the edge frequencies in the $i$th sketch, $S_i$, and $(m, n)$ is such an edge that is associated with $S_i$ for frequency maintenance and querying. Then, the *expected frequency* of $(m, n)$, denoted by $\bar{f}(m, n)$, when hashed into a cell of $S_i$ because of erroneous collisions, is determined by

$$\bar{f}(m,n) = (F(S_i) - f(m,n))/w_i$$

Therefore, the *expected relative error* of the edge $(m, n)$ is given by

$$\bar{e}(m,n) = \bar{f}(m,n)/f(m,n) = F(S_i)/(f(m,n) \cdot w_i) - 1/w_i$$

for any particular row in the sketch $S_i$. If the depth $d$ of the sketch $S_i$ is 1, the *overall relative error*, $E_i$, of the sketch $S_i$ is

$$E_i = \sum_{(m,n) \in S_i} \bar{e}(m,n) = \sum_{(m,n) \in S_i} \left( F(S_i)/(f(m,n) \cdot w_i) - 1/w_i \right) \quad (4)$$

Then the optimization problem of sketch partitioning can be formulated as follows:

**Problem** 1. *Partition the global sketch into $r$ localized sketches $S_1, \ldots, S_r$ based on the **edge set** of the data sample, with the objective to minimize $\sum_{i=1}^{r} E_i$, where $E_i$ is formulated in Equation (4).* □

Let us consider a simplification of this optimization problem in which we wish to construct $r = 2$ partitions. This is a difficult problem since it can be recast as a 0-1 integer program with a non-linear objective function [26]. There can be an exponential number of solutions to the problem and it is hard to determine the optimal one. We therefore seek an alternative solution. Our idea is to sort the edges in the global sketch in nondecreasing order of edge frequencies and consider those partitions containing edges in contiguously sorted order. The number of such partitions is equal to the number of edges, since we can choose the partition pivot at each possible position in the sorted order. However, the optimal partition pivot is picked at which the objective function in Equation (4) is minimized.

Unfortunately this solution is still not quite implementable, since we do not know the edge frequencies to begin with. Furthermore, the initial data sample is assumed to have a fairly small size compared to the actual graph stream. This means that the data sample cannot be reliably used to estimate the frequency of every edge in the graph stream. However, it can be effectively used to estimate the relative frequencies of vertices, as defined in Equation (2). Based on the property of local similarity of graph streams as described in Section 3.3, we alternatively use the frequency behavior of the *vertices* to perform the sketch partitioning. We denote the estimated relative frequency of a vertex $m$ by $\tilde{f}_v(m)$, and the estimated out degree of $m$ by $\tilde{d}(m)$. Both estimated values are derived from the data sample. Then the average frequency of the edges emanating from the vertex $m$ is determined by $\tilde{f}_v(m)/\tilde{d}(m)$. That is, we assume $\tilde{d}(m)$ edges emanate from the vertex $m$ with an average frequency of $\tilde{f}_v(m)/\tilde{d}(m)$. And the total estimated frequencies of the edges in the partitioned sketch $S_i$ $(1 \leq i \leq r)$, denoted as $\tilde{F}(S_i)$, can be expressed as

$$\tilde{F}(S_i) = \sum_{m \in S_i; m \in \mathcal{V}} \tilde{f}_v(m) \quad (5)$$

As a result, analogous to Equation (4), the overall relative error $E_i$ of the partitioned sketch $S_i$ with the use of vertex frequency-based statistics can be redefined as follows:

$$E_i = \sum_{m \in S_i} \frac{\tilde{d}(m) \cdot \tilde{F}(S_i)}{w_i \cdot \tilde{f}_v(m)/\tilde{d}(m)} - \sum_{m \in S_i} \tilde{d}(m)/w_i \quad (6)$$

Note that $\tilde{d}(m)$ in the numerator accounts for the fact that there are $O(\tilde{d}(m))$ edges emanating from the vertex $m$. The optimization problem of sketch partitioning is therefore transformed to the following form:

**Problem** 2. *Partition the global sketch into $r$ localized sketches $S_1, \ldots, S_r$ based on the **vertex set** of the data sample, with the objective to minimize $\sum_{i=1}^{r} E_i$, where $E_i$ is formulated in Equation (6).* □

As in the previous case, an approximate solution to this problem is to first sort the vertices in the data sample in order of average frequency, $\tilde{f}_v(m)/\tilde{d}(m)$, and then pick the partition pivot at which the objective function, as formulated in Problem 2, can be minimized.

By partitioning the global sketch based on the vertices, rather than the edges, we essentially create a set of localized sketches on different structural portions of the graph stream. The advantages of this vertex-based sketch partitioning approach are as follows. First of all, it intelligently relaxes the oracle assumption of Problem 1, and successfully transforms this hard optimization problem to a more tractable one, as described in Problem 2. Second, due to the sparsity of the data sample, the query estimation accuracy can be extremely low if the edge-based sketch partitioning approach is adopted. Instead, the vertex-based partitioning principle takes advantage of the local similarity property of the graph stream, which leads to a much more reliable and robust sketch partitioning method. Last but not least, the vertex-based partitioning information is compact and easy to compute [19]. This enables an efficient storage and maintenance of gSketch.

The analysis above suggests a natural way of constructing sketch partitions in a top-down recursive fashion as in decision trees. We call such a partitioning mechanism as a *partitioning tree*. At the first step, we have an initial root node $S$ representing the virtual global CountMin sketch with all the available space. The node is then split into two children $S_1$ and $S_2$, and the space allocated to either branch of $S$ is the same. This is done by evenly partitioning the width of the CountMin sketch corresponding to $S$ between two branches rooted with $S_1$ and $S_2$, respectively. In order to optimize such a partitioning of $S$ into $S_1$ and $S_2$, we need to minimize the objective function as expressed in Problem 2, which corresponds to the summation $E$ below:

$$\begin{aligned} E &= E_1 + E_2 \\ &= \sum_{m \in S_1} \frac{\tilde{d}(m) \cdot \tilde{F}(S_1)}{w_1 \cdot \tilde{f}_v(m)/\tilde{d}(m)} + \sum_{m \in S_2} \frac{\tilde{d}(m) \cdot \tilde{F}(S_2)}{w_1 \cdot \tilde{f}_v(m)/\tilde{d}(m)} \quad (7) \\ &\quad - \sum_{m \in S_1 \cup S_2} \tilde{d}(m)/w_1 \end{aligned}$$

Note the sketch widths of $S_1$ and $S_2$ are equal, i.e. $w_1 = w_2$. We therefore use $w_1$ uniformly throughout the expression in Equation (7). In order to further simplify the expression, we define an alternative expression $E'$ as

$$E' = E \cdot w_1 + \sum_{m \in S_1 \cup S_2} \tilde{d}(m) \quad (8)$$

It is obvious that $E$ is optimized whenever $E'$ is optimized. This is because $w_1$ is positive, and $\sum_{m \in S_1 \cup S_2} \tilde{d}(m)$ is a constant irrespective of how the partitioning of $S$ into $S_1$ and $S_2$ is performed. We simplify the value of $E'$ as follows:

$$E' = \sum_{m \in S_1} \frac{\tilde{d}(m) \cdot \tilde{F}(S_1)}{\tilde{f}_v(m)/\tilde{d}(m)} + \sum_{m \in S_2} \frac{\tilde{d}(m) \cdot \tilde{F}(S_2)}{\tilde{f}_v(m)/\tilde{d}(m)} \quad (9)$$

To this end, we evaluate the value of $E'$ over all possible partitions of $S$ in sorted order of $\tilde{f}_v(m)/\tilde{d}(m)$. Note there are as many choices of the partition pivot as the number of vertices in $S$, and we pick the one for which the value of $E'$ in Equation (9) is minimized. After $S$ is partitioned into $S_1$ and $S_2$, they themselves form the next pair of decision nodes for further partitioning consideration in the partitioning tree. For the internal nodes of the partitioning tree, we do not explicitly construct the corresponding sketches. Instead, we use them to maintain information for further sketch partitioning. This partitioning process is performed recursively until one of the following two termination criteria is met:

1. The width of a sketch $S$ at a given level is less than a particular threshold $w_0$, i.e., $S.width < w_0$;

2. The number of distinct edges being counted within a sketch $S$ is no greater than a constant factor $C$ of the sketch width, i.e., $\sum_{m \in S} \tilde{d}(m) \leq C \cdot S.width$.

In the first case, we do not further partition the sketch but build and materialize it explicitly, because the sketch of this kind is considered small enough and further partitioning will incur more collisions and therefore may hurt the final query estimation accuracy. The second termination case is determined by the following theorem:

THEOREM 1. *For a given sketch $S$ and a nonnegative constant $C$ $(0 < C < 1)$, s.t., $\sum_{m \in S} \tilde{d}(m) \leq C \cdot S.width$, the probability of any collision in a particular cell of $S$ can be bounded above by $C$.*

PROOF. We denote the hash function of a specific row of $S$ as $h(\cdot)$. Given two distinct edges $i$ and $j$, the probability of collision between $i$ and $j$ in a particular cell of $S$ can be determined as

$$\Pr(h(i) = h(j)) \leq 1/S.width$$

There are $\sum_{m \in S} \tilde{d}(m)$ distinct edges associated with the sketch $S$. By pairwise independence of the collision probability of distinct edges, we note the probability of any collision with the edge $i$ is

$$\sum_j \Pr(h(i) = h(j)) \leq \frac{\sum_{m \in S} \tilde{d}(m)}{S.width} \leq C$$

Therefore, the probability of any collision in a particular cell of $S$ is no greater than $C$. □

Intuitively, if the number of distinct edges within a sketch $S$ is small enough (bounded up by a constant factor of the width of $S$), the probability of collisions within $S$ will be small, and therefore $S$ can be directly used as a high quality localized sketch for query estimation without further partitioning. In practice, we further set the width of such sketches to the modest value of $\sum_{m \in S} \tilde{d}(m)$. It helps save extra space which can be allocated to other sketches, and improve the final query estimation accuracy. We note that even though the sketch partitioning is performed at every internal node of the partitioning tree, the sketches are physically constructed only at the leaves of the tree.

The sketch partitioning algorithm (with data sample only) is illustrated in Figure 2. We now give a detailed complexity analysis of the algorithm. In the partitioning tree, every internal node is split into two nodes for further inspection (Lines 6 − 7). In the worst case, the partitioning tree can be a complete binary tree with a height of $\log(w/w_0)$, and

**Algorithm** *Sketch-Partitioning* (Data Sample: $D$)
**begin**
1. Create a root node $S$ of the partitioning tree as an active node;
2. $S.width = w = \lceil e/\epsilon \rceil$;
3. $S.depth = d = \lceil \ln \frac{1}{\delta} \rceil$;
4. Create an active list $L$ containing $S$ only;
5. **while** $(L \neq \emptyset)$
   **begin**
6.   Partition an active node $S \in L$ based on $D$ into $S_1$ and $S_2$ by minimizing $E'$ in Equation (9);
7.   $S_1.width = S_2.width = S.width/2$;
8.   $L = L \setminus \{S\}$;
9.   **if** $(S_1.width \geq w_0)$ **and** $(\sum_{m \in S_1} \tilde{d}(m) > C \cdot S_1.width)$
10.    **then** $L = L \bigcup S_1$;
11.    **else** Construct the localized sketch $S_1$;
12.  **if** $(S_2.width \geq w_0)$ **and** $(\sum_{m \in S_2} \tilde{d}(m) > C \cdot S_2.width)$
13.    **then** $L = L \bigcup S_2$;
14.    **else** Construct the localized sketch $S_2$;
   **end**
**end**

Figure 2: Sketch Partitioning with Data Samples

the number of internal nodes in the partitioning tree can be at most $2^{\log(w/w_0)} - 1$, i.e., $(w/w_0 - 1)$, which is also the number of active nodes to be processed in $L$ (Line 5). For each internal node of the partitioning tree, we need to sort corresponding vertices and select the pivot at which the objective function can be minimized. The complexity of such operations is at most $O(|D|\log|D|)$. Therefore, the overall complexity of the algorithm is $O((w/w_0 - 1)|D|\log|D|)$.

## 4.2 Sketch Partitioning with Data and Workload Samples

In this section, we further assume that a query workload sample is available in addition to the data sample, and discuss how it can be exploited for more effective partitioning. In this scenario, it is possible to estimate the relative weights of different edge queries in the presence of the query workload sample. More specifically, the *relative weights of vertices* are estimated from the query workload sample and then incorporated into the sketch partitioning process. The relative weight of a vertex $n$ is the relative frequency of edges emanating from $n$ to be used in the querying process, and can be derived from the query workload sample. Let $\tilde{w}(n)$ be the relative weight of the vertex $n$ in the query workloads. In this scenario, the vertex based relative error, $E_i$, of the $i$th partitioned sketch, $S_i$, can be formulated as follows:

$$E_i = \sum_{n \in S_i} \frac{\tilde{w}(n) \cdot \tilde{F}(S_i)}{w_i \cdot \tilde{f}_v(n)/\tilde{d}(n)} - \sum_{n \in S_i} \tilde{w}(n)/w_i \quad (10)$$

This condition is similar to that formulated in Equation (6) for the data sample scenario. The difference is the term $\tilde{w}(n)$ in the numerator, which has been introduced in order to account for queries emanating from the vertex $n$.

During sketch partitioning, a given node $S$ in the partitioning tree is split into two nodes $S_1$ and $S_2$, such that the overall relative error is minimized. The objective function in this scenario can be formulated as follows:

$$E' = \sum_{n \in S_1} \frac{\tilde{w}(n) \cdot \tilde{F}(S_1)}{\tilde{f}_v(n)/\tilde{d}(n)} + \sum_{n \in S_2} \frac{\tilde{w}(n) \cdot \tilde{F}(S_2)}{\tilde{f}_v(n)/\tilde{d}(n)} \quad (11)$$

```
Algorithm Sketch-Partitioning (Data Sample: D;
            Workload Sample: W)
begin
1. Create a root node S of the partitioning tree as an
   active node;
2. S.width = w = ⌈e/ϵ⌉;
3. S.depth = d = ⌈ln 1/δ⌉;
4. Create an active list L containing S only;
5. while (L ≠ ∅)
   begin
6.     Partition an active node S ∈ L based on D and W
       into S_1 and S_2 by minimizing E' in Equation (11);
7.     S_1.width = S_2.width = S.width/2;
8.     L = L\{S};
9.     if (S_1.width ≥ w_0) and (∑_{m∈S_1} d̃(m) > C · S_1.width)
10.       then L = L ⋃ S_1;
11.       else Construct the localized sketch S_1;
12.    if (S_2.width ≥ w_0) and (∑_{m∈S_2} d̃(m) > C · S_2.width)
13.       then L = L ⋃ S_2;
14.       else Construct the localized sketch S_2;
   end
end
```

**Figure 3: Sketch Partitioning with Data and Workload Samples**

We sort the vertices in order of $\tilde{f}_v(n)/\tilde{w}(n)$ and perform the sketch partitioning at the pivot with which the objective function $E'$ in Equation (11) is minimized. The sketch partitioning algorithm for this scenario is shown in Figure 3. The major difference here is that we make use of both the data sample and the workload sample for sketch partitioning and the objective function is determined by Equation (11). Similarly, the worst-case time complexity of the algorithm in this scenario is $O((w/w_0 - 1)|D|\log|D|)$.

## 5. QUERY PROCESSING

Sketch partitioning is a pre-processing step to determine the association of vertices in the data sample to different partitioned localized sketches. More specifically, we maintain a hash structure $\mathcal{H} : \mathcal{V} \to S_i$, $1 \leq i \leq r$. For an edge $(m, n)$ in the graph stream, it can be hashed onto the localized sketch $\mathcal{H}(m) = S_i$ for frequency update and querying. Although this hash structure $\mathcal{H}$ is an extra overhead that needs to be stored along with the sketches, the cost is marginal compared to the immense advantages of sketch partitioning. Furthermore, we do not need to explicitly store the hierarchical structure of the partitioning tree. Only the partitioned sketches represented by the leaf nodes in the partitioning tree need to be physically stored for querying purposes.

After the off-line sketch partitioning phase, the resulting partitioned sketches are then populated with the massive graph stream in an online fashion and start supporting the query processing and estimation simultaneously. As the graph stream arrives, we use the hash structure $\mathcal{H}$ to associate incoming edges to the corresponding localized sketches, and update the edge frequencies in the corresponding sketch. During the querying phase, we analogously first determine the relevant partitioned sketch to which an edge query is associated. Once the sketch is identified by $\mathcal{H}$, the edge query can then be answered specifically by that sketch.

A special case is the one in which a particular vertex occurring in the graph stream does not occur in the original data sample. For edges which contain such vertices, a fixed portion of the original space is allocated as an *outlier partition* and a separate *outlier sketch* is constructed accordingly to count the frequencies of these edges. For the purpose of querying, those vertices which do not occur in any partition are resolved by this special outlier sketch. It is important to understand that the real graph streams are often considerably skew in vertex presence in data samples. The vertices involved in those edges with high frequency will typically be present in the sample as well. Therefore, it leaves only a small fraction of the overall *frequency* to be processed by the outlier sketch. Recall that the estimation error of the sketch-based methods is dependent upon the overall frequency of the items added to the sketch. Since most of the high frequency edges have already been skimmed off, the estimation results from the outlier sketch can still be accurate. Thus, even in the presence of new vertices in the graph stream, gSketch can achieve satisfactory query estimation results because of the removal of most of the high-frequency edges from the outlier sketch.

We note that the confidence intervals of the CountMin sketch method apply within each localized partition of gSketch. Since the number of edges assigned to each of the partitions is known in advance of query processing, it is possible to know the confidence of each particular query. Therefore, the confidence intervals of different queries are likely to be different depending upon the sketches that they are assigned to. On the average, the confidence intervals of different sketch partitions are likely to be similar to that of the global sketch with the same amount of space. However the exploitation of structural properties in gSketch leads to much better experimental behavior. We will present these experimental advantages in Section 6.

Users may sometimes be interested in dynamic queries over specific windows in time. For example, a user may be interested in the frequency behavior of edges in the past one month, one year and so on. In such cases, it makes sense to divide the time line into temporal intervals (or windows), and store the sketch statistics separately for each window. The partitioning in any particular window is performed by using a sample, which is constructed by reservoir sampling from the previous window in time. Queries over a specific time-interval can be resolved approximately by extrapolating from the sketch time windows which overlap most closely with the user-specified time window.

For the case of aggregate subgraph query, we first decompose the query into a bag of constituent edges and then sequentially process each edge as a separate edge query against the graph stream. Each such edge is first mapped to the appropriate sketch, and then is estimated by that sketch. After that, all estimated answers of the constituent edges are summarized by the aggregate function, $\Gamma(\cdot)$, as the final estimation result of the aggregate subgraph query.

## 6. EXPERIMENTAL RESULTS

In this section, we report our experimental findings on query estimation in graph streams. We compare our gSketch method with Global Sketch, which makes use of a global sketch for the entire graph stream in order for query estimation. Our experiments are evaluated in both scenarios characterized by the availability of data samples and query

199

workload samples. All our experiments are performed on an Intel PC with a 3.4 GHz CPU and 3.2GB main memory, running Window XP Professional SP3. All algorithms including gSketch and Global Sketch are implemented in C++.

### 6.1 Data Sets

We choose two real data sets and one synthetic data set in our experimental studies. Two of the data sets are publicly available, while one real data set is extracted from a large cooperate sensor network. The details of each data set are elaborated as follows.

**DBLP.** The DBLP database[2] contains scientific publications in the computer science domain, and we extract all conference papers ranging from 1956 to March 15th, 2008 for our experimental studies. There are 595,406 authors and 602,684 papers in total. We note that for a given paper, the authors are listed in a particular order as $a_1, a_2, \ldots, a_k$. An ordered author-pair $(a_i, a_j)$ is then generated if $1 \leq i < j \leq k$. There are 1,954,776 author-pairs in total, which are considered as streams of the underlying co-authorship graph, and are input to gSketch and Global Sketch in a chronological order.

**IP Attack Network.** Our second real data set is IP attack streaming data extracted from a corporate sensor network. The data set was initially collected from January 1st, 2007 to June 11th, 2007 comprising IP attack packet data from sensors. For each IP attack transaction, the attack type, time-stamp, sensor information, source IP address, target IP address and vulnerability status are recorded. We extract the source IP address and the target IP address of each IP attack packet to compose graph streams and select a time-frame from January 1st, 2007 to January 5th, 2007 as the time window of interest. This data set contains 3,781,471 edges in total.

**GTGraph.** The synthetic data set is generated by the well-known graph generator GTGraph[3]. A large network $G$ with power-law degree distributions and small-world characteristics is generated based on R-MAT model [9]. We choose default values of parameters during network generation, as suggested by the authors. The generated network contains $10^8$ vertices and $10^9$ edges, and the edges of $G$ are used as graph streams for experimental evaluation.

In order to verify the common characteristics of edge frequencies exhibited in graph streams, we denote the *global* variance of edge frequencies of the graph stream by $\sigma_G$. We further define the average *local* variance of edge frequencies on a vertex basis as $\sigma_V$. This is computed by determining the statistical variance of edge frequencies for the edges incident on each source vertex and averaging over different vertices. The *variance ratio*, $\sigma_G/\sigma_V$, for each of the three different data sets, DBLP, IP Attack Network, and GTGraph, is 3.674, 10.107, and 4.156, respectively. It is evident that the edge frequency variance on a vertex basis is consistently much smaller than the edge frequency variance of the whole graph stream. This also shows that there is significant skew in the frequency properties of graph streams, a fact we have considered in our sketch partitioning approach, gSketch.

### 6.2 Evaluation Methods

We evaluate different query estimation algorithms for both edge query and aggregate subgraph query. Edge queries are expressed as a set of edges drawn from the graph stream, whereas aggregate subgraph queries are expressed as a set of subgraphs whose aggregate frequency behavior is examined. Given an edge query set $Q_e = \{q_1, q_2, \ldots, q_k\}$, where $q_i$ is an edge in the graph stream, we consider two different accuracy measures for query estimation:

1. **Average Relative Error.** Given $q \in Q_e$, the relative error, $er(q)$, is defined as

$$er(q) = \frac{\tilde{f}(q) - f(q)}{f(q)} = \frac{\tilde{f}(q)}{f(q)} - 1 \quad (12)$$

Here, $\tilde{f}(q)$ and $f(q)$ are the estimated frequency and true frequency of $q$, respectively. The *average relative error of $Q_e$* is determined by averaging the relative errors over all queries of $Q_e$:

$$e(Q_e) = \frac{\sum_{i=1}^{k} er(q_i)}{k} \quad (13)$$

2. **Number of "Effective Queries".** Average relative error may become a biased measure if queries of $Q_e$ have significantly different frequencies. For example, if an edge with low frequency happens to collide with another edge with very high frequency in the sketch, this can lead to an extremely large value of average relative error. That is, a small number of such queries may dominate the overall average relative error of $Q_e$ in query estimation. We therefore propose another more robust measure, *number of effective queries*. The estimation of a query, $q$, is said to be "*effective*", if $er(q) \leq G_0$, where $G_0$ is a user-specified threshold. The idea here is that the estimated relative error of a query larger than $G_0$ may deviate too much from its true frequency, such that it is no longer considered as an *effective* estimation. We denote the number of effective queries estimated in $Q_e$ as $g(Q_e)$, and

$$g(Q_e) = |\{q|e(q) \leq G_0, q \in Q_e\}| \quad (14)$$

In all our experiments, we set $G_0 = 5$ by default, unless otherwise specified.

In addition to edge queries, we also consider aggregate subgraph queries. Given a subgraph query set $Q_g = \{g_1, g_2, \ldots, g_k\}$, the relative error of $g = \{e_1, \ldots, e_l\} \in Q_g$ is defined as

$$er(g) = \frac{\Gamma(\tilde{f}(e_1), \ldots, \tilde{f}(e_l))}{\Gamma(f(e_1), \ldots, f(e_l))} - 1 \quad (15)$$

In our experiments, we set $\Gamma(\cdot) = \text{SUM}(\cdot)$, so that the aggregate frequency behavior of a subgraph is summarized by adding up all the frequencies of the constituent edges of this graph. We define the query estimation measures (*average relative error*, $e(Q_g)$, and *number of "effective" queries*, $g(Q_g)$) for aggregate subgraph query in an analogous way as to edge query. As will be shown in the following sections, the query estimation results of aggregate subgraph query are similar to those of edge query. We therefore present the results of both kinds of queries in the DBLP data set only. For the other two data sets, we present the query estimation results only for edge queries. Besides query estimation accuracy, another important evaluation metric is the *efficiency* of sketch construction and query processing. The sketch construction time is therefore denoted as $T_c$, and the query processing time is denoted as $T_p$.

---

[2] http://www.informatik.uni-trier.de/~ley/db/
[3] http://www.cc.gatech.edu/~kamesh/GTgraph/index.htm

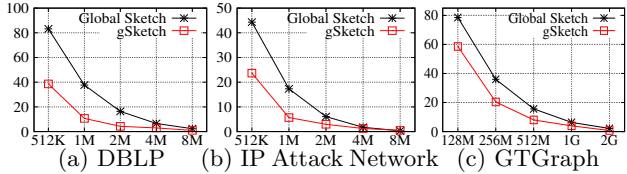

Figure 4: Average Relative Error of Edge Queries $Q_e$ w.r.t. Memory Size (Bytes)

### 6.3 Query Estimation with Data Samples

We first consider the scenario in which only the data sample is available for query estimation in different graph streams.

In the DBLP data set, a data sample with a size of 100,000 edges is generated by reservoir sampling [29] from the underlying co-authorship graph. The edge query set $Q_e$ comprises 10,000 edge queries, which are generated by uniform sampling. The subgraph query set $Q_g$ comprises 10,000 graphs, each of which is generated by first uniformly sampling vertices as seeds from the underlying co-authorship graph and then exploring the seeds' neighborhood by BFS traversal to include more vertices. At any given node during BFS traversal, the next edge to be explored is picked at random and each such subgraph contains 10 edges in total.

In the IP Attack Network data set, we select the IP pair streams from the first day (January 1st, 2007) as the data sample, which contains 445,422 edges. The query set $Q_e$ is generated from the original graph stream by uniform sampling and $|Q_e| = 10,000$.

For the synthetic GTGraph data set, we select 5,000,000 edges, i.e., 5% of the whole graph stream, as the data sample by reservoir sampling. The edge query set, $Q_e$, is selected from the graph stream as well by uniform sampling and $|Q_e| = 10,000$.

We first examine the query estimation accuracy of different algorithms w.r.t. the first evaluation metric, average relative error. The first set of results for the edge query sets, $Q_e$, is illustrated in Figure 4 across different data sets. It is evident that gSketch is consistently more accurate than Global Sketch at various memory footprints. This means the exploitation of underlying structural properties of graph streams is very helpful in boosting query estimation accuracy. When the available memory space is limited (less than 2M bytes in the DBLP and the IP Attack Network cases), the difference of query estimation accuracy of the two algorithms is very large. For example, in the DBLP data set, gSketch can achieve $2 - 8$ times better estimation results than Global Sketch. In the more interesting space-constrained scenarios, this difference in estimation accuracy becomes very significant, and it suggests that gSketch can be used in extremely space-constrained devices, such as sensors, for effective query estimation. This also suggests that the accuracy difference of two methods will still exist when the size of the underlying graph increases, as shown in the GTGraph case (Figure 4(c)). When the graph stream becomes large with $10^9$ edges, gSketch still outperforms Global Sketch even when the available memory becomes large up to 2G bytes. This is an important result, because most typical graph stream applications such as social networks continue to become more and more massive over time. When the available memory becomes large, the difference of estimation accuracy between gSketch and Global Sketch reduces, because theoretically both methods can estimate the queries accurately if given infinitely large memory. However, gSketch still outperforms Global Sketch in such cases.

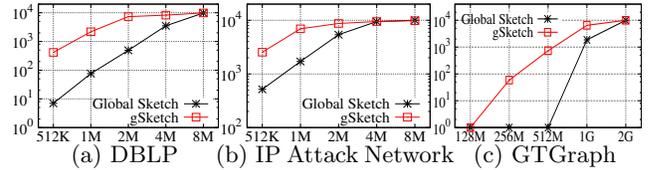

Figure 5: Number of Effective Queries for Edge Queries $Q_e$ w.r.t. Memory Size (Bytes)

We then evaluate the query estimation accuracy in terms of the number of effective queries estimated in the edge query set $Q_e$. The estimation results across different data sets are illustrated in Figure 5 (log-scaled). Again gSketch outperforms Global Sketch by as much as one or two orders of magnitude. Interestingly, in the synthetic GTGraph data set, Global Sketch simply fails to estimate any edge queries "effectively" when the available space is less than 512M bytes. However, gSketch can achieve much better estimation results. It suggests that when the graph stream becomes massive and the underlying graph exhibits significant global heterogeneity and local similarity properties, Global Sketch is no longer a feasible solution, while gSketch continues to retain its effectiveness.

Next, we evaluate the estimation accuracy of aggregate subgraph queries in the DBLP graph stream. As illustrated in Figures 6, gSketch consistently outperforms Global Sketch in terms of both average relative error and number of effective queries estimated. The experimental results also verify that gSketch is much more effective than Global Sketch for estimating *both* edge queries and aggregate subgraph queries.

### 6.4 Query Estimation with Data and Workload Samples

In this scenario, we evaluate the query estimation accuracy of different algorithms when both the data sample and the query workload sample are available from graph streams. We generate the query workload samples from different graph streams as follows. For the DBLP data set, we generate a query workload sample with 400,000 edges by sampling (without replacement) the graph stream which follows the Zipf distribution. The Zipf-based sampling is parameterized by a skewness factor $\alpha$. The larger the value of $\alpha$, the more skewed the query workload sample. It is worth mentioning that a vertex $m$ that exists in the data sample may not necessarily appear in the query workload sample. For such a case, we use the Laplace smoothing [22] to avoid $\tilde{w}(m)$, the estimated relative weight of $m$, to be zero. Edge and subgraph queries are generated in a similar way by Zipf-based sampling and the sizes of both the edge query set and the subgraph query set are 10,000. In the IP Attack Network, we construct a query workload sample with 800,000 edges by Zipf sampling. and the edge query set is generated with $|Q_e| = 10,000$. For the synthetic GTGraph data set, we generate the query workload sample by Zipf sampling from the original graph, which contains $5,000,000$ edges, and the edge query set is generated with $|Q_e| = 10,000$.

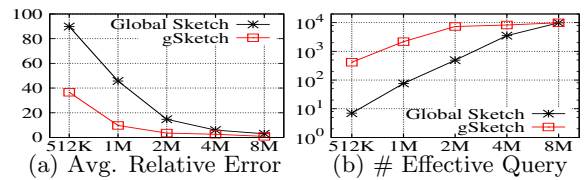

Figure 6: Query Estimation Accuracy of Graph Queries $Q_g$ w.r.t. Memory Size (Bytes) in DBLP Data Set

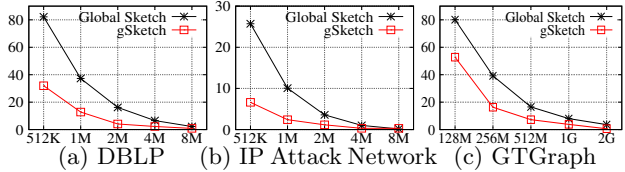

Figure 7: Average Relative Error of Edge Queries $Q_e$ w.r.t. Memory Size(Bytes) (Zipf Skewness $\alpha = 1.5$)

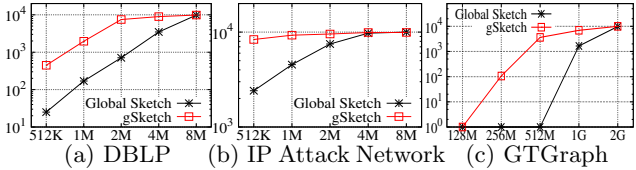

Figure 8: Number of Effective Queries for Edge Queries $Q_e$ w.r.t. Memory Size (Bytes) (Zipf Skewness $\alpha = 1.5$)

In our first experiment, we fix $\alpha = 1.5$ for Zipf sampling during the generation of workload samples and queries. We then examine the query estimation accuracy in the presence of both data and query workload samples. The average relative error and number of effective queries for edge queries on all data sets are reported in Figure 7 and Figure 8, respectively. In this scenario, gSketch still outperforms Global Sketch consistently at various memory footprints across different data sets. More interestingly, the estimation accuracy is higher than that for the case when the data samples are available only (as shown in Figure 4 and Figure 5) because of the involvement of the query workload samples. This accuracy gain is observed under both evaluation metrics, because we are leveraging a greater amount of querying information in sketch partitioning. It may sometimes happen that frequently occurring edges in the query workload sample may not be present as frequently in the data sample. Such edges can be estimated far more accurately in this scenario and further contribute to the improved accuracy.

We then evaluate the query estimation accuracy by varying the value of the skewness factor, $\alpha$, to generate a set of query workloads. The available memory here is fixed with 2M bytes for the DBLP data set and the IP Attack Network data set, and 1G bytes for the GTgraph data sets, throughout this experiment. The average relative error of query estimation accuracy is illustrated in Figure 10 across different data sets. It is evident that with the increase of $\alpha$, gSketch can achieve better estimation accuracy with a decreasing trend of average relative errors, because larger values of $\alpha$ lead to more skewness in the query workload. Such skewness are accounted for in the sketch partitioning process and help improve the final query estimation results. On the contrary, Global Sketch is unable to use such query workload information, and therefore there is no such accuracy improvement. The number of effective queries for the different methods is illustrated in Figure 11 across different data sets. Similar to the previous case in which average relative error is adopted as the evaluation metric, gSketch achieves better estimation accuracy with an increasing trend of the number of effective queries estimated, when $\alpha$ varies from 1.2 to 2.0. This accuracy improvement results from a better usage of the workload samples during the sketch partitioning.

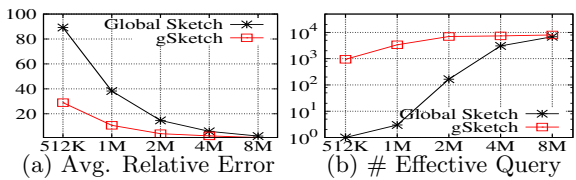

Figure 9: Query Estimation Accuracy of Graph Queries $Q_g$ w.r.t. Memory Size (Bytes) in DBLP Data Set ($\alpha = 1.5$)

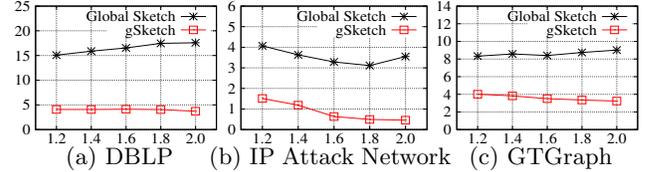

Figure 10: Average Relative Error of Edge Queries $Q_e$ w.r.t. Zipf Sampling Skewness $\alpha$

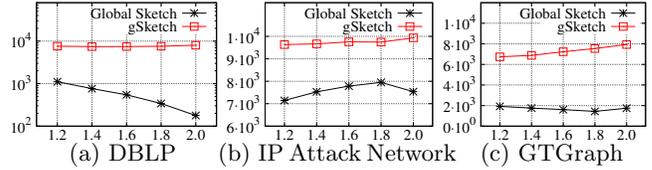

Figure 11: Number of Effective Queries for Edge Queries $Q_e$ w.r.t. Zipf Sampling Skewness $\alpha$

Similar experimental evaluations are performed for aggregate subgraph queries on the DBLP data set. In Figure 9, the query estimation accuracy is reported with the sampling factor $\alpha = 1.5$. In Figure 12, $\alpha$ varies from 1.2 to 2.0 and the query estimation accuracy is reported if the available memory is 2M bytes. Analogous to edge queries, aggregate subgraph queries can be more effectively estimated with the use of query workload information in gSketch. On the other hand, Global Sketch performs even worse due to the frequency heterogeneity of the constituent edges in subgraph queries.

### 6.5 Efficiency Results

Besides query estimation accuracy, the time used for sketch construction, $T_c$, and the time used for query processing, $T_p$, are evaluated as well. Figure 13 illustrates $T_c$ for gSketch

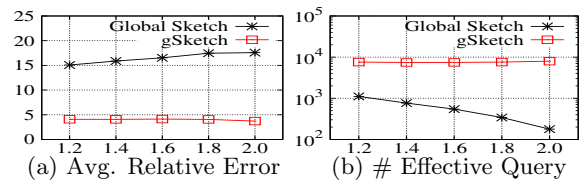

Figure 12: Query Estimation Accuracy of Graph Queries $Q_g$ w.r.t. Zipf Sampling Skewness $\alpha$ in DBLP Data Set

202

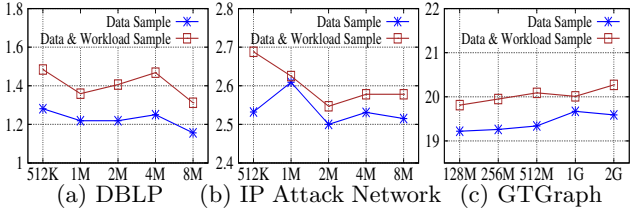

Figure 13: Sketch Construction Time $T_c$ (Seconds) w.r.t. Memory Size (Bytes)

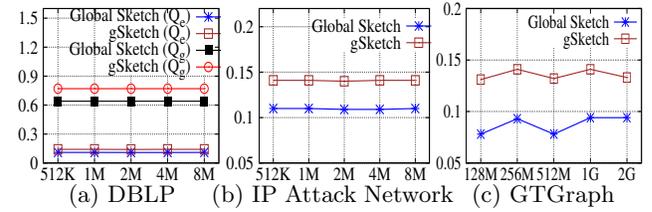

Figure 14: Query Processing Time $T_p$ (Seconds) w.r.t. Memory Size (Bytes)

across different data sets for the two different scenarios. For the scenario in which both data and query workload sample are available, $T_c$ is slightly larger than that when only the data sample is available. In both scenarios, however, $T_c$ is within tens of seconds even for the largest graph stream, GTGraph, which contains $10^9$ edges. Therefore, gSketch can be rapidly constructed and effectively deployed from a practical point of view.

We further examine the query processing time, $T_p$, for both gSketch and Global Sketch across different data sets, as shown in Figure 14. For the DBLP data set, we measure the query processing time for both edge queries and aggregate subgraph queries. For the other two data sets, we present the query processing time only for edge queries. First of all, $T_p$ for gSketch is insensitive to the allocated memory, as long as the sketch structure can be built and allocated successfully in the sketch construction phase. It is evident that each edge query can be estimated within 0.2 seconds for all different data sets, while the response time for an aggregate subgraph query is within 0.8 seconds for the DBLP data set. Therefore, gSketch can be used in real time on different graphs streams. Compared with Global Sketch, gSketch takes additional time to determine the relevant sketch partition a query belongs to. However, such time is negligible and does not hurt the practicality of gSketch. On the other hand, the enormous advantages of the sketch-partitioning philosophy definitely outweigh such cost.

## 6.6 Effect of New Vertices

To this end, we assume the underlying graph model of a graph stream is static. While in reality, such a graph is dynamically changing with new vertices and corresponding edges created all the time, thus forming an ever-growing massive network. As these newly added vertices are not in the sample data, their corresponding edges will be uniformly assigned to the outlier sketch for query estimation. A key question is *how these vertices affect the query estimation process*. We will demonstrate that gSketch is robust to the presence of such new vertices upon the underlying graph.

In order to illustrate this point, we present the estimation accuracy only for the edge queries answered by the outlier sketch, and compare it to the accuracy of all the edge queries answered by gSketch. Table 1 illustrates such query estimation accuracy in terms of average relative error on the GT-Graph data set (For other data sets and scenarios, we have similar results). It is evident that the outlier sketch does not significantly degrade our query estimation accuracy. The main reason is the outlier sketch has already been skimmed of the high frequency edges, which could potentially cause collisions for estimation inaccuracy. The estimation error in the outlier sketch is therefore not significantly higher than other partitioned sketches. This also suggests that to the presence of new vertices and edges on the underlying graph, gSketch is still a robust solution to estimating queries on the graph stream.

## 6.7 Discussion on Experimental Studies

After extensive studies of gSketch and Global Sketch in different experimental settings on various data sets, the following conclusions can be made:

1. When space is considered a critical factor, gSketch consistently achieves better query estimation accuracy on large graph streams in terms of both average relative error and number of effective queries estimated. Furthermore, the importance of the space limitation increases with the domain size of the underlying graph.

2. By exploiting both data sample and query workload sample, gSketch can achieve better query estimation accuracy than that achieved with only the data sample.

3. When both data samples and query workload samples are available, gSketch will benefit if the samples are skewed. The more skewed the query workload sample, the better query estimation accuracy gSketch may achieve.

4. For gSketch, the time spent for sketch partitioning and construction is marginal. Furthermore, query processing can be performed very fast and the time is relatively invariant to the allocated space.

5. To the presence of new vertices and their corresponding edges on the underlying graph of a graph stream, gSketch is still a robust solution for query estimation, as long as the estimation error of the outlier sketch is not significantly higher than that of gSketch.

## 7. CONCLUSIONS

The problem of querying graph streams is very challenging because of the high stream speed and the massive universe of distinct edges. In this paper, we designed an effective sketch-partitioning algorithm, gSketch, for query estimation over massive graph streams. We made use of the special structural properties of graph streams to help devise a sketch partitioning solution in order to improve the query estimation accuracy. We tested our approach on a number of real and synthetic data sets and illustrated the advantages of our sketch partitioning algorithm, gSketch, over a global sketching scheme, Global Sketch. gSketch has proven to be significantly more effective, and sometimes provides query estimation accuracy better than Global Sketch by an order of magnitude.

|  |  | Memory | Size | | | |
| --- | --- | --- | --- | --- | --- | --- |
|  |  | 128M | 256M | 512M | 1G | 2G |
| **Average** | gSketch | 58.5968 | 20.381 | 8.0068 | 3.9345 | 0.7257 |
| **relative error** | Outlier sketch | 58.5971 | 20.392 | 8.0081 | 3.9557 | 0.7837 |

Table 1: Avg. Relative Error of gSketch and Outlier Sketch in GTGraph Data Set

In future work, we will study methods for resolving more complex queries such as those involving the computation of complex functions of edge frequencies in a subgraph query. We will also examine the use of sketch-based methods for resolving structural queries. Finally, we will investigate how such sketch-based methods can be potentially designed for dynamic analysis, which may not require any samples for constructing the underlying synopsis.

## Acknowledgement

The research was sponsored in part by the Army Research Laboratory and was accomplished under Cooperative Agreement Number W911NF-09-2-0053. The views and conclusions contained in this document are those of the author and should not be interpreted as representing the official policies, either expressed or implied, of the Army Research Laboratory or the U.S. Government. The U.S. Government is authorized to reproduce and distribute reprints for Government purposes notwithstanding any copyright notice hereon.

## 8. REFERENCES

[1] C. C. Aggarwal. *Data Streams: Models and Algorithms*. Springer, Inc., 2006.
[2] C. C. Aggarwal. *Social Network Data Analytics*. Springer Inc., 2011.
[3] C. C. Aggarwal, Y. Li, P. S. Yu, and R. Jin. On dense pattern mining in graph streams. *Proc. VLDB Endow.*, 3:975–984, 2010.
[4] C. C. Aggarwal and H. Wang. *Managing and Mining Graph Data*. Springer Inc., 2010.
[5] N. Alon, Y. Matias, and M. Szegedy. The space complexity of approximating the frequency moments. In *STOC*, pages 20–29, Philadelphia, PA, USA, 1996.
[6] Z. Bar-Yossef, R. Kumar, and D. Sivakumar. Reductions in streaming algorithms, with an application to counting triangles in graphs. In *SODA*, pages 623–632, San Francisco, CA, USA, 2002.
[7] P. Boldi and S. Vigna. The webgraph framework I: compression techniques. In *WWW*, pages 595–602, New York, NY, USA, 2004.
[8] L. S. Buriol, G. Frahling, S. Leonardi, A. Marchetti-Spaccamela, and C. Sohler. Counting triangles in data streams. In *PODS*, pages 253–262, Chicago, IL, USA, 2006.
[9] D. Chakrabarti, Y. Zhan, and C. Faloutsos. R-MAT: A recursive model for graph mining. In *SDM*, pages 442–446, Lake Buena Visa, FL, USA, 2004.
[10] F. Chierichetti, R. Kumar, S. Lattanzi, M. Mitzenmacher, A. Panconesi, and P. Raghavan. On compressing social networks. In *KDD*, pages 219–228, Paris, France, 2009.
[11] E. Cohen and H. Kaplan. Tighter estimation using bottom k sketches. *Proc. VLDB Endow.*, 1:213–224, 2008.
[12] D. J. Cook and L. B. Holder. *Mining Graph Data*. John Wiley & Sons, 2006.
[13] G. Cormode and M. Hadjieleftheriou. Finding frequent items in data streams. *Proc. VLDB Endow.*, 1:1530–1541, 2008.
[14] G. Cormode and S. Muthukrishnan. An improved data stream summary: the count-min sketch and its applications. *J. Algorithms*, 55:58–75, 2005.
[15] G. Cormode and S. Muthukrishnan. Space efficient mining of multigraph streams. In *PODS*, pages 271–282, Baltimore, Maryland, USA, 2005.
[16] A. Das Sarma, S. Gollapudi, and R. Panigrahy. Estimating pagerank on graph streams. In *PODS*, pages 69–78, Vancouver, BC, Canada, 2008.
[17] A. Dobra, M. Garofalakis, J. Gehrke, and R. Rastogi. Processing complex aggregate queries over data streams. In *SIGMOD*, pages 61–72, Madison, WI, USA, 2002.
[18] M. Faloutsos, P. Faloutsos, and C. Faloutsos. On power-law relationships of the internet topology. *SIGCOMM Comp. Comm. Rev.*, 29:251–262, 1999.
[19] J. Feigenbaum, S. Kannan, A. McGregor, S. Suri, and J. Zhang. On graph problems in a semi-streaming model. *Theor. Comput. Sci.*, 348:207–216, 2005.
[20] S. Ganguly and B. Saha. On estimating path aggregates over streaming graphs. In *ISAAC*, pages 163–172, Kolkata, India, 2006.
[21] M. R. Henzinger, P. Raghavan, and S. Rajagopalan. Computing on data streams. *External memory algorithms*, pages 107–118, 1999.
[22] D. Jurafsky and J. H. Martin. *Speech and Language Processing*. Prentice Hall, 2008.
[23] G. S. Manku and R. Motwani. Approximate frequency counts over data streams. In *VLDB*, pages 346–357, Hong Kong, China, 2002.
[24] A. McGregor. Graph mining on streams. *Encyclopedia of Database Systems*, pages 1271–1275, 2009.
[25] S. Muthukrishnan. Data streams: algorithms and applications. *Found. Trends Theor. Comput. Sci.*, 1:117–236, 2005.
[26] G. L. Nemhauser and L. A. Wolsey. *Integer and combinatorial optimization*. Wiley-Interscience, 1988.
[27] M. E. J. Newman. *Network: An Introduction*. Oxford University Press, 2010.
[28] S. Raghavan and H. Garcia-Molina. Representing web graphs. In *ICDE*, pages 405–416, Atlanta, GA, USA, 2003.
[29] J. S. Vitter. Random sampling with a reservoir. *ACM Trans. Math. Softw.*, 11:37–57, 1985.